# Topological hyperbolic lattices

Sunkyu Yu[†], Xianji Piao, and Namkyoo Park[*]

Photonic Systems Laboratory, Dept. of Electrical and Computer Engineering, Seoul National University, Seoul 08826, Korea

**Abstract**

Non-Euclidean geometry, discovered by negating Euclid's parallel postulate, has been of considerable interest in mathematics and related fields for the description of geographical coordinates, Internet infrastructures, and the general theory of relativity. Notably, an infinite number of regular tessellations in hyperbolic geometry—hyperbolic lattices—can extend Euclidean Bravais lattices and the consequent band theory to non-Euclidean geometry. Here we demonstrate topological phenomena in hyperbolic geometry, exploring how the quantized curvature and edge dominance of the geometry affect topological phases. We report a recipe for the construction of a Euclidean photonic platform that inherits the topological band properties of a hyperbolic lattice under a uniform, pseudospin-dependent magnetic field, realizing a non-Euclidean analogue of the quantum spin Hall effect. For hyperbolic lattices with different quantized curvatures, we examine the topological protection of helical edge states and generalize Hofstadter's butterfly, showing the unique spectral sensitivity of topological immunity in highly curved hyperbolic planes. Our approach is applicable to general non-Euclidean geometry and enables the exploitation of infinite lattice degrees of freedom for band theory.

Band theory in condensed-matter physics and photonics, which provides a general picture of the classification of a phase of matter, has been connected to the concept of topology[1-4]. The discovery of a topologically nontrivial state in a crystalline insulator, having a knotted property of a wavefunction in reciprocal space, has revealed a new phase of electronic[3], photonic[4], and acoustic[5] matter: the topological insulator. This topological phase offers the immunity of electronic conductance[1,3] or light transport[2,6-8] against disorder, occurring only on the boundary of the matter through topologically protected edge states.

A fundamental approach for generalizing band theory and its topological extension is to rethink the traditional assumptions regarding energy-momentum dispersion relations, such as static, Hermitian, and periodic conditions. In photonics, dynamical lattices can derive an effective magnetic field for photons, realizing one-way edge states in space-time Floquet bands[9]. Non-Hermitian photonics introduces novel band degeneracies[10] and topological phenomena[11] in complex-valued band structures. Various studies have extended band theory to disordered systems, achieving perfect bandgaps in disordered structures[12,13] and topological invariants in amorphous media[14-16].

Meanwhile, most of the cornerstones in band theory have employed Euclidean geometry, because Bloch's theorem is well defined for a crystal, which corresponds to the uniform tessellation of Euclidean geometry. However, significant lattice degrees of freedom are overlooked in wave phenomena confined to Bloch-type Euclidean geometry because of the restricted inter-elemental relationships from a finite number of uniform Euclidean tessellations. For example, when considering a lattice with congruent unit cells, only six, four, or three nearest-neighbour interactions are allowed in two-dimensional (2D) Euclidean lattices[17-19] due to three types (triangular, square, and hexagonal) of regular tiling. In this context, access to non-Euclidean geometry has been of considerable interest for employing more degrees of freedom in wave manipulations, as reported in metamaterials[20] and transformation optics[21]. Recently, in circuit quantum electrodynamics, the extension of band structures to hyperbolic geometry was demonstrated[22], which not only revealed unique flat bands with spectral isolation but also allowed for a significant extension of a phase of matter through non-Euclidean lattice configurations.

Here, we demonstrate topological phases of matter in hyperbolic geometry. We develop a 2D Euclidean photonic platform that inherits the band properties of a "hyperbolic lattice": the lattice obtained from the regular tiling of the hyperbolic plane. In addition to the enhanced band degeneracy in topologically trivial phases, we reveal topologically protected helical edge states with strong but curvature-dependent immunity against disorder as a non-Euclidean photonic analogue of the quantum spin Hall effect (QSHE)[1-4]. The Hofstadter butterfly[23], which represents the fractal energy spectrum of 2D electrons in a uniform magnetic field, is also generalized to photonic hyperbolic geometry. Our theoretical result is a step towards the generalization of topological wave phenomena to non-Euclidean geometries, which allows access to an infinite number of lattice degrees of freedom.

**Hyperbolic lattices**

Among three types of homogeneous 2D geometries—elliptic (positive curvature), Euclidean (zero curvature), and hyperbolic (negative curvature) planes—only Euclidean and hyperbolic planes belong to an infinite plane[17,18], allowing for infinite lattice structures. Thus, we focus on topological phenomena in 2D hyperbolic lattices[22] as a non-Euclidean extension of topological lattices. Hyperbolic lattices are obtained by generalizing Euclidean Bravais lattices to regular tiling: a tessellation of a plane using a single type of regular polygon. Each vertex of a polygon then corresponds to a lattice element, while the polygon edge between vertices generalizes the Bravais vector that describes nearest-neighbour interactions. To visualize the wrinkled geometry of the hyperbolic plane[24] and achieve 2D effective Euclidean platforms for hyperbolic lattices[22] as discussed later, we employ the Poincaré disk model[17,18]: the projection of a hyperboloid onto the unit disk (Fig. 1a-c, bottom).

The negative curvature of the hyperbolic plane differentiates hyperbolic lattices from their Euclidean counterpart. First, a polygon in the hyperbolic plane has a smaller sum of internal angles than that of the Euclidean counterpart, permitting denser contact of the polygons at the vertices (Fig. 1a-c). For example, while only four squares can be contacted at a vertex of a Euclidean square lattice, in principle, *any* number of squares greater than four can be contacted at a vertex in hyperbolic lattices, because the square has a curvature-dependent internal angle less than $\pi/2$. Such infinite lattice degrees of freedom can be expressed by the Schläfli symbol $\{p,q\}$, the contact of $q$ $p$-sided regular polygons around each vertex[17,18], as $\{4,q\geq 5\}$, $\{5,q\geq 4\}$, and $\{6,q\geq 4\}$ for square, pentagon, and hexagon unit cells, respectively. Each regular tiling is achieved by recursively adding neighbouring polygons starting from a seed polygon (Fig. 1d), which eventually fills the unit disk at the infinite lattice limit. This recursive process implies tree-like geometric properties of hyperbolic lattices.

Infinite lattice degrees of freedom in the hyperbolic plane instead restrict the allowed unit cell area of a hyperbolic lattice. Contrary to freely tunable areas of Euclidean polygons, the polygon area in the hyperbolic plane is restricted by the curvature of the plane and the sum of the internal angles[17,18]. The Gaussian curvature $K$ of the hyperbolic lattice $\{p,q\}$ is then uniquely determined as (Supplementary Note S1)

$$K = -\frac{p\pi}{A_{\text{poly}}}\left(1 - \frac{2}{p} - \frac{2}{q}\right), \tag{1}$$

where $A_{\text{poly}}$ is the area of the lattice unit polygon. For the same $A_{\text{poly}}$, the curvature $K$ of a hyperbolic lattice has a "quantized" value defined by the lattice configuration $\{p,q\}$.

Because of the crinkling shapes of the hyperbolic plane[17,24], the ideal realization of 2D hyperbolic lattices is extremely difficult, requiring very complicated 3D structures. To resolve this issue, we employ graph modelling of wave systems[22,25], which assigns graph vertices to the system elements and graph edges to identical interactions between the elements. For the 2D Euclidean realization of a hyperbolic lattice, we employ the Poincaré disk as a graph network. The intrinsic properties of a wave system then depend only on its graph topology, leading to identical wave properties for Euclidean realizations of the Poincaré disk graph and its continuously deformed graphs (Fig. 1e).

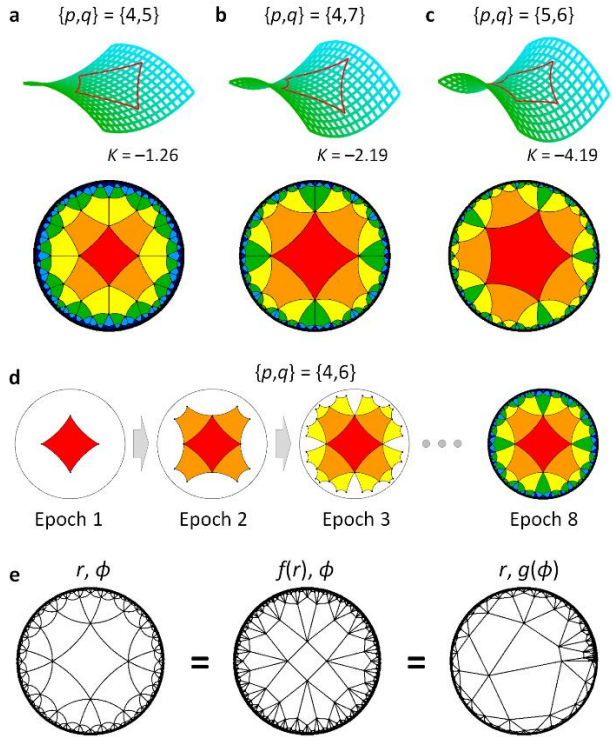

**Fig. 1. Hyperbolic lattices with quantized curvatures. a-c,** (top) Local view of the saddle-shaped surface of the hyperbolic plane and (bottom) Poincaré disks of hyperbolic lattices for different Gaussian curvatures $K$: **a,** $K = -1.26$ for $\{p,q\} = \{4,5\}$, **b,** $K = -2.19$ for $\{p,q\} = \{4,7\}$, and **c,** $\{p,q\} = K = -4.19$ for $\{5,6\}$. $A_{\text{poly}} = 1$ in the calculation of $K$. Red polygons on the surfaces have the same area. Each colour in the Poincaré disk represents a different epoch in generating the lattice. **d,** Recursive generation of a hyperbolic lattice. **e,** A graph view of the hyperbolic lattice. For the original vertex positions in the polar coordinate $(r,\phi)$, each graph obtained from the deformation examples $f(r) = r^2$ and $g(\phi) = \phi^2 / (2\pi)$ results in the same wave properties as the original graph.

**Photonic hyperbolic lattices**

For the 2D Euclidean realization of the Poincaré disk graph, we employ photonic coupled-resonator platforms, assigning each lattice site (or graph vertex) to the resonator that supports the pseudo-spins for clockwise ($\sigma = +1$) and counter-clockwise ($\sigma = -1$) circulations. The identical connection between lattice sites (or graph edge) is realized by the indirect coupling between nearest-neighbour resonators through a zero-field waveguide loop (Supplementary Note S2), which has a $(4m + 1)\pi$ phase evolution[6,27] (Fig. 2a-c). The strength of the indirect coupling is independent of the real-space distance between the resonators and is determined solely by the evanescent coupling $\kappa$ between the waveguide loop and the resonators: the same indirect coupling strength in Fig. 2a,b. By introducing the additional phase difference between the upper $(+\varphi)$ and lower $(-\varphi)$ arms (Fig. 2c), the waveguide loop leads to the gauge field $\varphi$ having a different sign for each pseudo-spin[2]. The gauge field distribution then determines the effective magnetic field for pseudo-spins[2,26].

The platform in Fig. 2a-c enables a real-space construction of the Poincaré disk graph (Fig. 2d), which inherits wave properties of the corresponding hyperbolic lattice. The structure is governed by the photonic tight-binding Hamiltonian (Supplementary Note S2)

$$H = \sum_{m,\sigma} \omega_m a_{m\sigma}^\dagger a_{m\sigma} + t \sum_{\langle m,n \rangle, \sigma} \left( e^{-i\sigma\varphi_{mn}} a_{m\sigma}^\dagger a_{n\sigma} + h.c. \right), \quad (2)$$

where $\omega_m$ is the resonance frequency of the $m^{th}$ resonator and is set to be constant as $\omega_m = \omega_0$, $a_{m\sigma}^\dagger$ (or $a_{m\sigma}$) is the creation (or annihilation) operator for the $\sigma$ pseudo-spin at the $m^{th}$ lattice site, $t = \kappa^2/2$ is the indirect coupling strength between the nearest-neighbour sites, $\varphi_{mn}$ is the additionally acquired phase from the $n^{th}$ to $m^{th}$ sites ($\varphi_{mn} = -\varphi_{nm}$), the nearest-neighbour pair $\langle m,n \rangle$ is determined by the edge of the Poincaré disk graph, and $h.c.$ denotes the Hermitian conjugate. The practical realization of the Poincaré disk graph is restricted by the unique geometric nature of hyperbolic lattices: the minimum distance between the resonators, which exhibits scale invariance to the resonator number (Supplementary Note S3).

In analysing the Poincaré disk graph and its wave structure, we consider spin-degenerate finite structures. First, due to the lack of commutative translation groups and Bravais vectors in hyperbolic geometry, Bloch's theorem cannot be applied to hyperbolic lattices[22]. We instead apply numerical diagonalization to the Hamiltonian $H$ for finite but large hyperbolic lattices. Second, because pseudo-spin modes experience the same magnetic field strength with the opposite sign, the consequent band structures of both pseudo-spins are identical when spin mixing, such as the Zeeman effect, is absent[2]. Assuming this spin degeneracy, we focus on the $\sigma = +1$ pseudo-spin.

Figure 2e shows topologically trivial ($\varphi_{mn} = 0$) eigenenergy spectra, comparing the Euclidean $\{4,4\}$ and hyperbolic $\{4, q \geq 5\}$ square lattices. Because of a large number of nearest-neighbour interactions and the consequent increase in the number of structural symmetries, the hyperbolic lattices lead to enhanced spectral degeneracy when compared to the Euclidean lattice (pink regions in Fig. 2e). Hyperbolic square lattices thus provide significantly extended flat bands, as observed in hyperbolic Kagome lattices in circuit quantum electrodynamics[22].

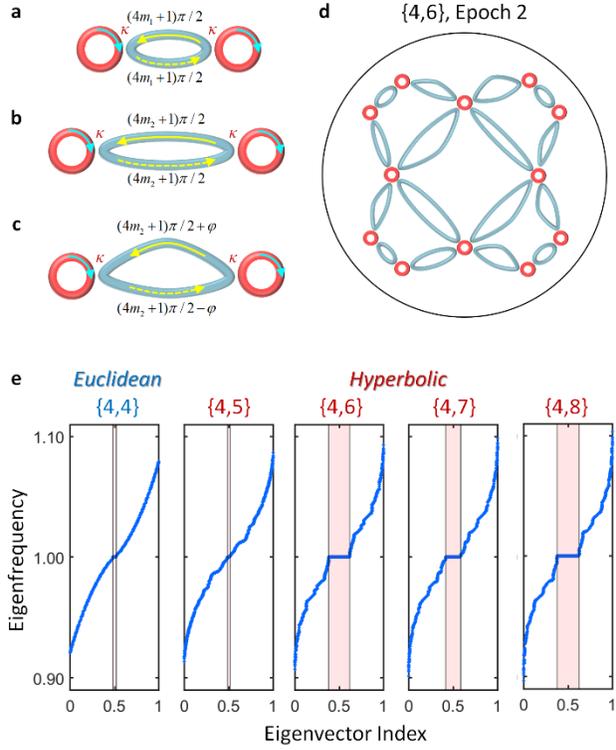

**Fig. 2. Photonic hyperbolic lattices with enhanced degeneracy. a-c,** Schematics for waveguide-based indirect coupling. The arrows denote wave circulations inside the resonators and waveguide loops. While the indirect coupling strength $t = \kappa^2/2$ is identical for **a-c,** the additional phase $\pm\varphi$ in **c** leads to the gauge field. **d,** Photonic hyperbolic lattice {4,6}. **e,** Topologically trivial eigenenergy spectra normalized by $\omega_0$ for the 25×25 Euclidean square lattice and different types of hyperbolic square lattices for the 6th epoch generation. The pink shaded regions represent the flat bands. $t = \omega_0 / 50$.

**Hyperbolic quantum spin Hall effect**

We now investigate topologically nontrivial states in hyperbolic geometry by implementing the photonic QSHE[2,4] in hyperbolic square lattices {4,$q\geq$5}, which we call the "hyperbolic QSHE". The uniqueness of the hyperbolic QSHE stems from non-commutative translation groups in hyperbolic geometry[18,22], which prohibit the natural counterpart of Bravais lattices. When constructing a uniform magnetic field in topological Euclidean lattices, a "Landau gauge"—the linearly increasing gauge field along one of the Bravais vectors—is usually employed[1-4,9,23,26]. However, this Landau gauge cannot be naturally implemented in hyperbolic geometry due to the absence of Bravais vectors. To achieve topological hyperbolic lattices, we instead propose the systematic design of a "hyperbolic gauge" for the hyperbolic QSHE (Fig. 3a,b), which exploits the tree-like, recursive generation process of hyperbolic lattices (Fig. 1d). We note that in evaluating the magnetic field and flux from the assigned gauge field, the areas of all of the squares $A_{poly}$ are effectively the same for the entire class of $\{4,q\geq5\}$ ($A_{poly} = 1$ for simplicity), because the indirect coupling strength $t$ is set to be the same. The identical $A_{poly}$ also uniquely determines the quantized curvature $K$ according to Eq. (1).

Consider the target magnetic flux $\theta$ through each unit polygon (or plaquette) of a hyperbolic lattice. First, we assign gauge fields equally to each edge of the seed polygon (blue solid arrows in Fig. 3a). For the polygons in the next epoch (Fig. 3b), after evaluating the predefined gauge fields (blue dashed arrows), we calculate the deficient gauge for the target flux $\theta$ in each polygon and then assign the necessary average gauge to each undefined polygon edge (green solid arrows). The recursive process eventually leads to the gauge configuration that achieves a uniform magnetic field ($B = \theta$ for $A_{poly} = 1$) through the entire polygons (Fig. 3c). This systematic procedure is applicable to *any* geometry under an arbitrary magnetic field distribution, including elliptic geometry, different polygons, and non-uniform magnetic fields.

When analysing topological wave phenomena in hyperbolic lattices, we cannot apply the well-known reciprocal-space formulation of the Chern number due to the absence of Bravais vectors. Although the real-space formulation of the Chern number has been studied for the estimation of the topological nature in aperiodic systems[14-16], we instead employ an empirical quantity $C_{edge}^{(n)}$ defined by[27]

$$C_{\text{edge}}^{(n)} = \frac{\sum_{r \in \Lambda_{\text{edge}}} |\psi_r^{(n)}|^2}{\sum_{r=1}^{N} |\psi_r^{(n)}|^2}, \qquad (3)$$

where $r$ denotes each lattice element, $N$ is the total number of elements, $\psi_r^{(n)}$ is the field amplitude of the $n^{\text{th}}$ eigenstate at the $r^{\text{th}}$ element, and $\Lambda_{\text{edge}}$ is the set of the edge elements of the system. $C_{\text{edge}}^{(n)}$ measures the spatial energy concentration at the system edge and has been applied to quantify the topological phase in disordered systems[27].

Figure 3d-f shows $C_{\text{edge}}^{(n)}$ in the Euclidean lattice (Fig. 3d) and hyperbolic lattices with different curvatures (Fig. 3e,f) at each eigenfrequency $\omega$ for the magnetic field $B = 0$. Figure 3d presents the spectrum analogous to Hofstadter's butterfly[23], where the regions of high $C_{\text{edge}}^{(n)}$ (red-to-yellow) depict topologically protected helical edge states in the QSHE[1-5]. In contrast, hyperbolic lattices generally lead to a much higher $C_{\text{edge}}^{(n)}$ than the Euclidean lattice (Fig. 3e,f with the colour range: $0.9 \leq C_{\text{edge}}^{(n)} \leq 1.0$), showing stronger energy confinement on the system edge. This "edge-dominant" behaviour is explained by the tree-like geometric nature of hyperbolic lattices: more edge elements in a more curved geometry, analogous to more leaves on a tree with more branches (Supplementary Note S4). This geometrical origin—a high edge-to-bulk ratio of hyperbolic lattices—also clarifies that a high $C_{\text{edge}}^{(n)}$ does not guarantee "topological protection" in contrast to Euclidean disordered systems[27] due to the possible existence of topologically trivial edge states.

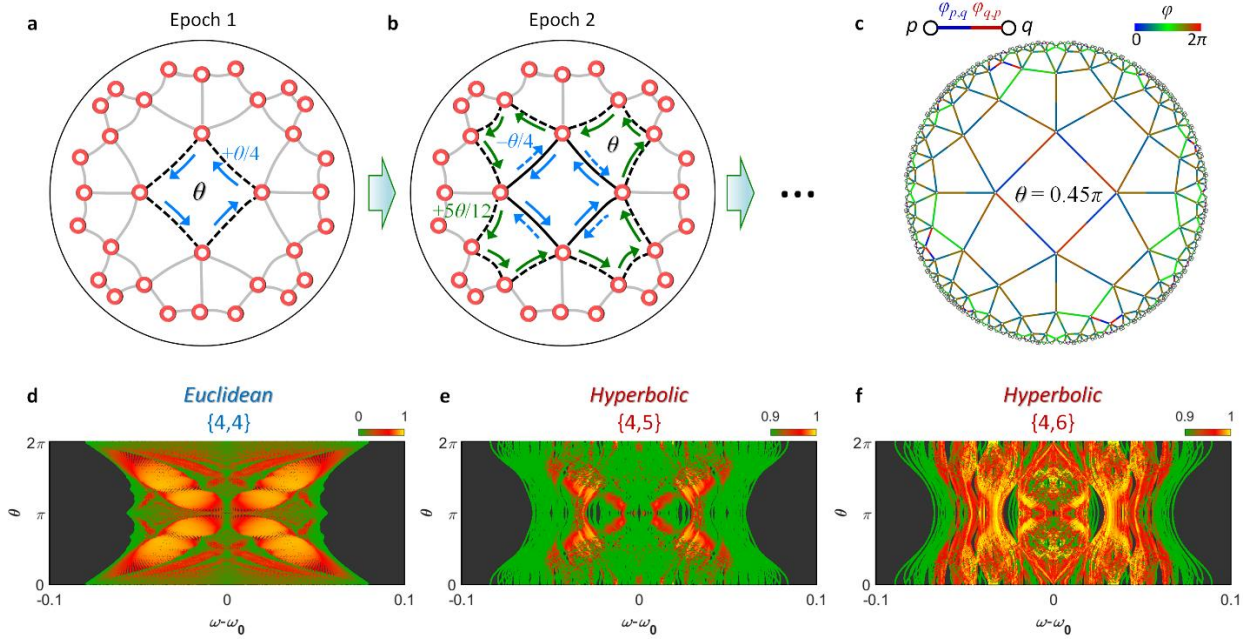

**Fig. 3. Realization of the hyperbolic quantum spin Hall effect. a-c,** Design of a "hyperbolic gauge" for a uniform magnetic field: **a,** seed polygon, **b,** polygons in the next epoch, and **c,** the final result. In **a** and **b,** the blue (or green) arrows denote the gauge field assigned in the first (or second) epoch. In **c,** the gauge field $\varphi_{q,p}$ at each polygon edge represents the gauge field from the $p^{\text{th}}$ to $q^{\text{th}}$ nearest-neighbour element. **d-f,** $C_{\text{edge}}^{(n)}$ as a function of the eigenfrequency $\omega$ and the uniform magnetic field $B = \theta$ for $A_{\text{poly}} = 1$: **d,** Euclidean lattice {4,4} with $K = 0$, and **e,f,** hyperbolic lattices of **e,** {4,5} with $K = -1.26$ and **f,** {4,6} with $K = -2.09$. $\omega_0 = 1$ for simplicity. All other parameters are the same as those in Fig. 2.

**Topological protection and Hofstadter's hyperbolic butterflies**

To extract "topologically protected" states from a high density of (topologically trivial or nontrivial) edge states, we examine the scattering from hyperbolic QSHE systems, following previous works on the photonic QSHE[2,4]. We employ input and output waveguides that are evanescently coupled to the selected elements at the system boundary (Fig. 4a,b) and then evaluate the transmission against disorder. To preserve the spin degeneracy, we focus on spin-mixing-free diagonal disorder[2], which may originate from imperfections in the radii or refractive indices of optical resonators. In Eq. (2), diagonal disorder is globally introduced by assigning random perturbations to the resonance frequency of each element, as $\omega_m = \omega_0 + \text{unif}[-\Delta,+\Delta]$, where unif[$a,b$] is a uniform random distribution between $a$ and $b$ and $\Delta$ is the maximum perturbation strength.

Figure 4c shows the transmission spectrum at a magnetic field of $B = \theta = 0.9\pi$ for an ensemble of 50 realizations of disorder. We observe spectral bands with high transmission (points **a** and **b**), which correspond to topologically protected helical edge states[5-8,10] with backward or forward rotations around the system (Fig. 4a,b). Furthermore, as shown in the small standard deviation of the transmission, the topologically protected states maintain their high transmission for different realizations of disorder, successfully achieving immunity to disorder.

To identify this topological protection, we introduce another empirical parameter that measures the disorder-immune transmission: $C_{\text{immune}}(\omega, \theta) = \text{E}[t(\omega, \theta)]$, where E[…] denotes the expectation value and $t(\omega, \theta)$ is the power transmission spectrum for each realization at a magnetic field of $B = \theta$. We compare $C_{\text{immune}}(\omega, \theta)$ in the Euclidean lattice (Fig. 4d) and the hyperbolic lattices with different curvatures $K$ (Fig. 4e,f), achieving the hyperbolic counterpart of the Hofstadter butterfly. While the Euclidean lattice again presents a $C_{\text{immune}}$ map analogous to the conventional Hofstadter butterfly[23,28] (Fig. 4d), the $C_{\text{immune}}$ maps for hyperbolic lattices (Fig. 4e,f) display a significant discrepancy from the $C_{\text{edge}}^{(n)}$ maps in Fig. 3e,f. This distinction demonstrates that the empirical parameter $C_{\text{immune}}$ successfully extracts "disorder-immune", topologically protected edge states from the entire (topologically trivial or nontrivial) edge states.

Despite the continuous Schläfli symbols of Euclidean {4,4} and hyperbolic {4,$q\geq5$} lattices, the patterns of Hofstadter's butterflies are separately classified according to the lattice geometry: Euclidean (Fig. 4d) and hyperbolic (Fig. 4e,f) lattices. This evident distinction emphasizes the topological difference between Euclidean and hyperbolic geometries[17,18], in that Euclidean geometry cannot be obtained from hyperbolic geometry through continuous deformations, and vice versa, as shown in the absence of Bravais vectors in hyperbolic geometry. Therefore, the classical interpretation of a fractal energy spectrum in Hofstadter's butterfly, focusing on the commensurability between the lattice period and magnetic length[23,28] in Harper's equation[29], cannot be straightforwardly extended to hyperbolic counterparts. Instead, the origin of the narrower spectral-magnetic ($\omega$-$\theta$) bandwidths of the topological protection in hyperbolic lattices (Fig. 4e,f) is explained by their edge-dominant geometry (Supplementary Note 4). A high density of topologically trivial edge states from the edge-dominant geometry (Fig. 3d-f) hinders the exclusive excitation of topologically nontrivial edge states.

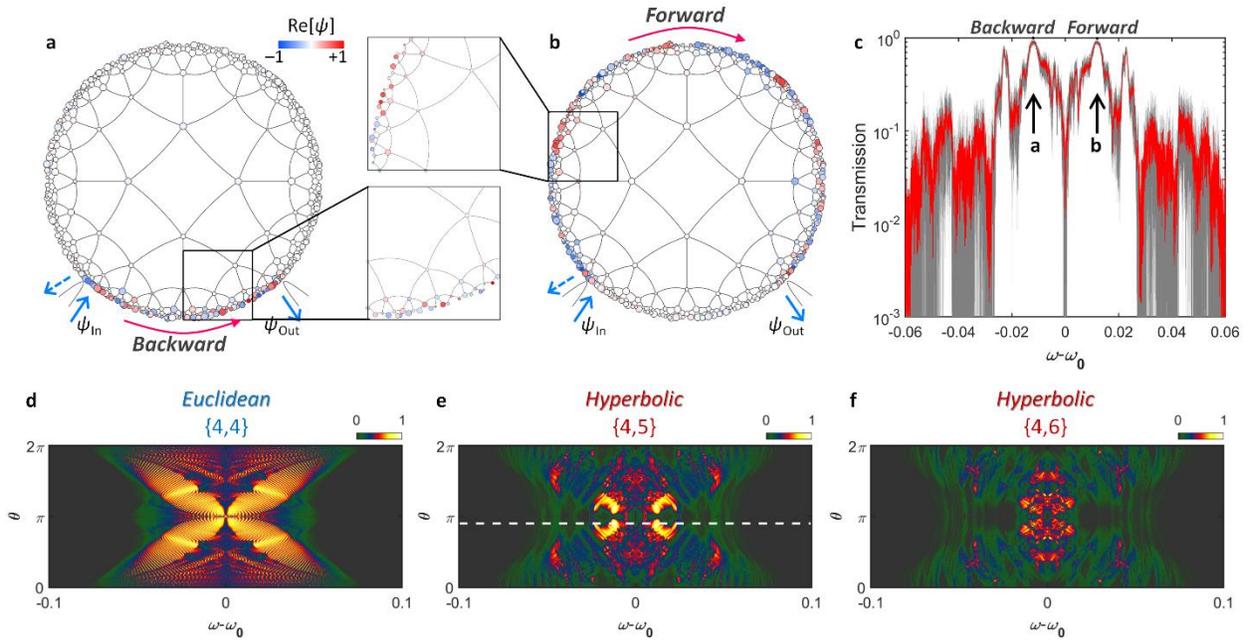

**Fig. 4. Topologically protected edge states and Hofstadter's "hyperbolic" butterflies**. **a,b,** Transmission through topologically protected helical edge states in the hyperbolic lattice {4,5}: **a,** backward ($t$ = 99.50% at $\omega$ = –0.012$\omega_0$) and **b,** forward ($t$ = 98.51% at $\omega$ = +0.012$\omega_0$) transmission, for $B = \theta = 0.9\pi$ (white dashed line in **e**). Different radii of the elements represent diagonal disorder: the resonators with different resonance frequencies $\omega_m$. **c,** Transmission

spectrum for an ensemble of 50 realizations of disorder (the red line is the average and the grey regions represent the standard deviation). **d-f,** $C_{\text{immune}}(\omega, \theta)$ obtained from 50 realizations of diagonal disorder: **d,** Euclidean lattice {4,4}, and **e,f,** hyperbolic lattices of **e,** {4,5} with $K = -1.26$ and **f,** {4,6} with $K = -2.09$. The maximum perturbation strength for diagonal disorder is $\Delta = \omega_0 / 200$ for all cases. The coupling between the selected boundary elements and the input or output waveguide is $0.08\omega_0$. All other parameters are the same as those in Fig. 3.

**Discussion**

We have demonstrated topological wave properties in hyperbolic geometry. By employing a systematic gauge field design, we achieved a hyperbolic lattice under a uniform magnetic field—a topological hyperbolic lattice—which leads to the hyperbolic counterpart of the QSHE. Using two empirical parameters that measure the edge confinement ($C_{\text{edge}}^{(n)}$) and disorder immunity ($C_{\text{immune}}$), we classified a high density of edge states in hyperbolic lattices in terms of topological protection. With the spectrally sensitive topological immunity in highly curved hyperbolic lattices, we expect novel frequency-selective photonic devices with error robustness. Our approach will also inspire the generalization of topological hyperbolic lattices in other classical or quantum systems involving acoustics[5] or cold atoms[26].

As observed in distinct patterns of Hofstadter's Euclidean and hyperbolic butterflies, hyperbolic geometry is topologically distinguished from Euclidean geometry. This topological uniqueness is emphasized with the geometrical nature of hyperbolic lattices shown in our results: the scale invariance and high edge-to-bulk ratio, which are conserved under the control of the $q$ parameter in {$p,q$} (or the quantized curvature $K$). We expect the engineering of disorder[12,13,15,16,25,27] exploiting the topological uniqueness of hyperbolic geometry, which will open up a new route for scale-free networks in material design, as already implied in Internet infrastructures[30]. A 3D real-space construction of hyperbolic lattices using origami design[24] may also be of considerable interest.

**Acknowledgements**

We acknowledge financial support from the National Research Foundation of Korea (NRF) through the Global Frontier Program (S.Y., X.P., N.P.: 2014M3A6B3063708), the Basic Science Research Program (S.Y.: 2016R1A6A3A04009723), and the Korea Research Fellowship Program (X.P., N.P.: 2016H1D3A1938069), all funded by the Korean government.

**Author contributions**

S.Y. conceived the idea presented in the manuscript. S.Y. and X.P. developed the numerical analysis for hyperbolic lattices. N.P. encouraged S.Y. and X.P. to investigate non-Euclidean geometry for waves while supervising the findings of this work. All authors discussed the results and contributed to the final manuscript.

**Competing interests**

The authors have no conflicts of interest to declare.

**Correspondence and requests for materials** should be addressed to N.P. (nkpark@snu.ac.kr) or S.Y. (skyu.photon@gmail.com).

**Supplementary Information for "Topological hyperbolic lattices"**

Sunkyu Yu[†], Xianji Piao, and Namkyoo Park[*]

Photonic Systems Laboratory, Department of Electrical and Computer Engineering, Seoul National University, Seoul 08826, Korea

*E-mail address for correspondence: nkpark@snu.ac.kr (N.P.); skyu.photon@gmail.com (S.Y.)

**Note S1. Quantized curvatures of hyperbolic lattices**

**Note S2. Derivation of the photonic tight-binding Hamiltonian**

**Note S3. Scale-invariant distance between resonators**

**Note S4. Edge dominance in hyperbolic lattices**

**Note S1. Quantized curvatures of hyperbolic lattices**

In this note, we show the relationship between the curvature of a hyperbolic lattice and the geometric properties of the lattice unit polygon. We start by dividing the unit polygon of the lattice $\{p,q\}$ into $p$ identical triangles (Fig. S1a). Each triangle then has the sum of internal angles $\xi_{sum} = 2\pi(1/p + 1/q)$. Because the area of a triangle $A_{tri}$ in the hyperbolic plane is determined by $\xi_{sum}$ and the Gaussian curvature $K$ as $A_{tri} = (\pi - \xi_{sum}) / (-K)$ [1,2], the unit polygon area is $A_{poly} = pA_{tri} = -(p\pi/K)\cdot(1 - 2/p - 2/q)$. The curvature of the hyperbolic lattice $\{p,q\}$ for the given polygon area $A_{poly}$ is then uniquely determined by Eq. (1) in the main text.

Figure S1b shows the Gaussian curvature $K$ of hyperbolic lattices, which have different lattice configurations. When the unit polygon area $A_{poly}$ is fixed, only quantized curvature values are allowed for the lattice in the hyperbolic plane (circles in Fig. S1b). The increase in the number of nearest-neighbour interactions $q$ or the number of polygon vertices $p$ leads to more curved lattices having higher values of $|K|$. Because all of the polygon edges are realized with identical waveguide-based indirect coupling, which is independent of the real-space length (Fig. 2a-c in the main text), all of the results in the main text adopt the same value of $A_{poly}$. We set $A_{poly} = 1$ for simplicity throughout this paper.

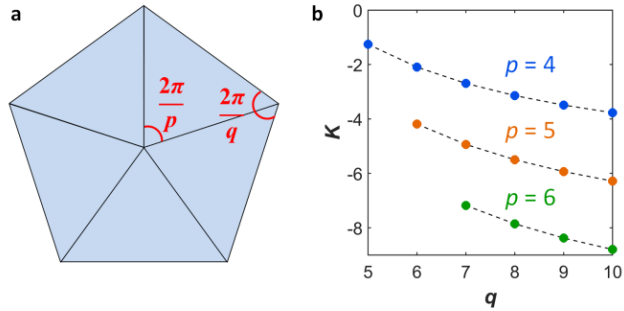

**Fig. S1. The relationship between quantized curvatures and lattice configurations. a,** A schematic of a lattice unit polygon for estimating the curvature of the lattice $\{p,q\}$. **b,** Gaussian curvature $K$ for different sets of $\{p,q\}$ with $A_{poly} = 1$.

**Note S2. Derivation of the photonic tight-binding Hamiltonian**

Consider the $m^{th}$ and $n^{th}$ nearest-neighbour optical resonators in a hyperbolic lattice. The resonators are indirectly coupled through the waveguide loop (Fig. S2), and the waveguide loop and each resonator are evanescently coupled with the coupling coefficient $\kappa$. The coupled mode equation for a single pseudo-spin mode in each resonator $\psi_m$ and $\psi_n$ ($\sigma = +1$) is then [3,4]

$$\frac{d}{dt}\begin{bmatrix}\psi_m \\ \psi_n\end{bmatrix} = \begin{bmatrix} i\omega_m - \frac{1}{2\tau} & 0 \\ 0 & i\omega_n - \frac{1}{2\tau} \end{bmatrix}\begin{bmatrix}\psi_m \\ \psi_n\end{bmatrix} + \sqrt{\frac{1}{\tau}}\begin{bmatrix}\mu_{mI} \\ \mu_{nI}\end{bmatrix}, \quad (S1)$$

$$\begin{bmatrix}\mu_{mO} \\ \mu_{nO}\end{bmatrix} = \begin{bmatrix}\mu_{mI} \\ \mu_{nI}\end{bmatrix} - \begin{bmatrix}\kappa & 0 \\ 0 & \kappa\end{bmatrix}\begin{bmatrix}\psi_m \\ \psi_n\end{bmatrix}, \quad (S2)$$

$$\begin{bmatrix}\mu_{mI} \\ \mu_{nI}\end{bmatrix} = \begin{bmatrix}0 & e^{-i\Phi_{mn}} \\ e^{-i\Phi_{nm}} & 0\end{bmatrix}\begin{bmatrix}\mu_{mO} \\ \mu_{nO}\end{bmatrix}, \quad (S3)$$

where $\tau = 1/\kappa^2$ is the lifetime of a single pseudo-spin mode inside each resonator, $\mu_{mI}$, $\mu_{mO}$, $\mu_{nI}$, and $\mu_{nO}$ are the field amplitudes at each position of the waveguide loop, as shown in Fig. S2, and $\Phi_{mn}$ is the phase evolution from the $n^{th}$ to $m^{th}$ resonators along the waveguide.

Using Eqs. (S2) and (S3), $\mu_{mI}$ and $\mu_{nI}$ can be expressed with the resonator fields as

$$\begin{bmatrix}\mu_{mI} \\ \mu_{nI}\end{bmatrix} = \frac{\kappa}{1-e^{i(\Phi_{mn}+\Phi_{nm})}}\begin{bmatrix}1 & e^{i\Phi_{nm}} \\ e^{i\Phi_{mn}} & 1\end{bmatrix}\begin{bmatrix}\psi_m \\ \psi_n\end{bmatrix}. \quad (S4)$$

Substituting Eq. (S4) into Eq. (S1), we obtain the following coupled mode equation:

$$\frac{d}{dt}\begin{bmatrix}\psi_m \\ \psi_n\end{bmatrix} = \begin{bmatrix} i\omega_m - \frac{1}{2\tau} & 0 \\ 0 & i\omega_n - \frac{1}{2\tau} \end{bmatrix}\begin{bmatrix}\psi_m \\ \psi_n\end{bmatrix} + \frac{1/\tau}{1-e^{i(\Phi_{mn}+\Phi_{nm})}}\begin{bmatrix}1 & e^{i\Phi_{nm}} \\ e^{i\Phi_{mn}} & 1\end{bmatrix}\begin{bmatrix}\psi_m \\ \psi_n\end{bmatrix}. \quad (S5)$$

To confine light inside the resonators rather than the waveguide loop, the phase evolution along the loop leads to destructive interference, while the evolution along each arm results in constructive interference with the resonator field. These conditions are satisfied with the phase evolutions $\Phi_{mn} = 2q_{mn}\pi + \pi/2 + \varphi_{mn}$ and $\Phi_{nm} = 2q_{nm}\pi + \pi/2 + \varphi_{nm}$ ($q_{mn}$ and $q_{nm}$ are integers), where $\varphi_{mn} + \varphi_{nm} = 2p\pi$ ($p$ is an integer). For a similar length of each arm of the waveguide loop, we set $q_{mn} = q_{nm} = q$ and $\varphi_{mn} = -\varphi_{nm}$. Equation (S5) then becomes

$$\frac{d}{dt}\begin{bmatrix}\psi_m \\ \psi_n\end{bmatrix} = \begin{bmatrix} i\omega_m & i\frac{1}{2\tau}e^{-i\varphi_{mn}} \\ i\frac{1}{2\tau}e^{+i\varphi_{mn}} & i\omega_n \end{bmatrix}\begin{bmatrix}\psi_m \\ \psi_n\end{bmatrix}. \tag{S6}$$

Considering both pseudo-spin modes ($\sigma = \pm 1$) and introducing the creation (or annihilation) operator for the $\sigma$ pseudo-spin mode of the $m^{th}$ resonator $a_{m\sigma}^{\dagger}$ (or $a_{m\sigma}$), Eq. (S6) derives the following photonic tight-binding Hamiltonian [4]

$$H = \sum_{\sigma}\left[\omega_m a_{m\sigma}^{\dagger}a_{m\sigma} + \omega_n a_{n\sigma}^{\dagger}a_{n\sigma} + t\left(e^{-i\sigma\varphi_{mn}}a_{m\sigma}^{\dagger}a_{n\sigma} + h.c.\right)\right], \tag{S7}$$

where $t = 1/(2\tau) = \kappa^2/2$ is the indirect coupling strength between the $m^{th}$ and $n^{th}$ resonators and $h.c.$ denotes the Hermitian conjugate. Notably, the sign of the pseudo-spin $\sigma$ determines the sign of the gauge field. Applying Eq. (S7) to all of the nearest-neighbour resonator pairs in a hyperbolic lattice, we obtain Eq. (2) in the main text.

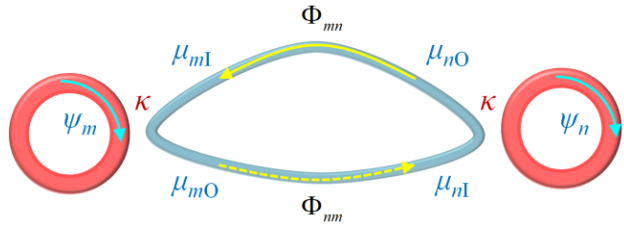

**Fig. S2. Coupled mode theory for the photonic tight-binding Hamiltonian.** A schematic for the coupled mode theory that describes the indirect coupling between optical resonators through the waveguide loop.

**Note S3. Scale-invariant distance between resonators**

Figure S3 shows the minimum distance between lattice resonators for different types of lattice configurations $\{4, q \geq 5\}$ and the generation epoch. In sharp contrast to the constant distance between nearest-neighbour resonators in Euclidean lattices, the increase in the hyperbolic lattice size leads to a significant decrease in the minimum resonator-to-resonator distance. We note that this distance is the characteristic length for a practical realization of hyperbolic lattices. Remarkably, the relationship between the minimum distance and the number of lattice resonators exhibits "scale invariance" following the power law [5-7], as demonstrated by the linear relationship in the log-log plot (Fig. S3). This result shows the close connection between hyperbolic geometry and scale-free networks such as Internet infrastructures [8].

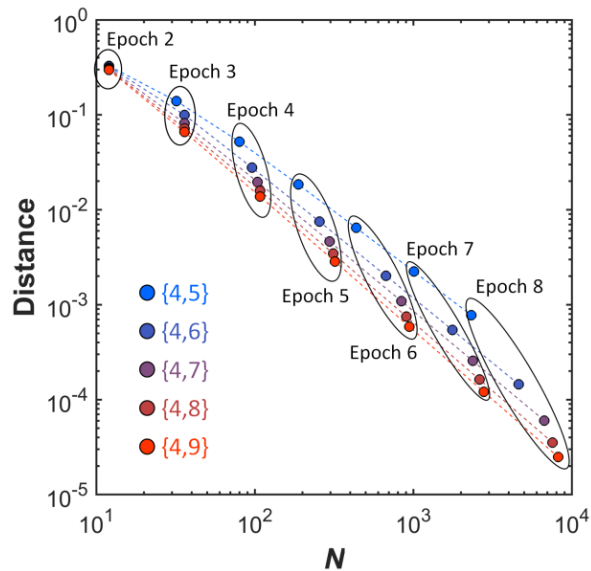

**Fig. S3. Scale invariance of the minimum resonator-to-resonator distance with respect to the number of resonators in a hyperbolic lattice.** The log-log plot shows the relationship between the number of lattice resonators and the minimum distance between the resonators. The distance is normalized by the radius of the Poincaré disk. The symbols represent each epoch for the generation of hyperbolic lattices $\{4, q \geq 5\}$.

**Note S4. Edge dominance in hyperbolic lattices**

The recursive generation in Fig. 1d in the main text, which implies the tree-like geometric nature of hyperbolic lattices, shows that more "branches" (or more polygons meeting at the vertex: a larger $q$ in $\{p,q\}$) will lead to more "leaves" (or more elements at the system edge). Figure S4 demonstrates this prediction, presenting the portion of the edge elements relative to the entire elements $N_{bnd} / N$, where $N$ is the total number of elements and $N_{bnd}$ is the number of elements at the system boundary. The hyperbolic lattices $\{4, q \geq 5\}$ show the evident edge-dominant geometric nature ($N_{bnd} / N \geq 0.8$), and the edge-to-bulk ratio becomes higher for a higher value of $q$.

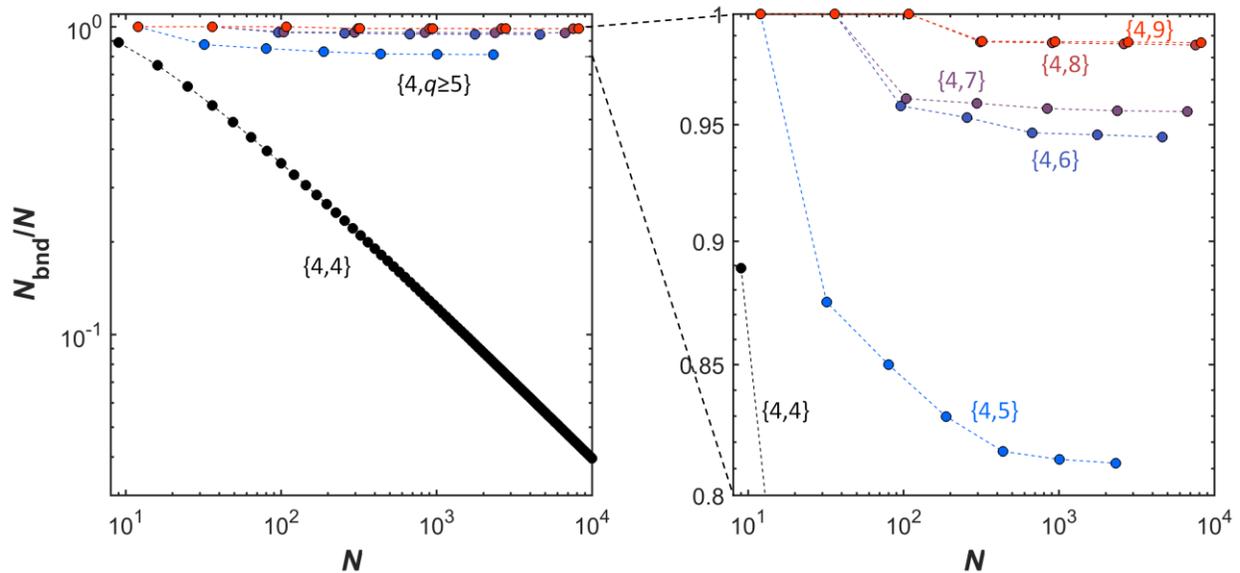

**Fig. S4. Edge dominance in hyperbolic lattices.** For Euclidean $\{4,4\}$ and hyperbolic $\{4, q \geq 5\}$ lattices, the edge-to-bulk ratio is presented by the portion of edge elements $N_{bnd} / N$ as a function of the lattice size $N$. The right figure is an enlarged plot of the left figure. Each point for the hyperbolic lattices is obtained in different generation epochs (epochs 2 to 8).